\documentclass[%
 reprint,
 onecolumn,
nofootinbib,
 amsmath,amssymb,
 aps,
prd,
]{revtex4-2}

\usepackage[colorlinks=true,linkcolor=blue,citecolor=blue,urlcolor=blue]{hyperref}
\usepackage{orcidlink}

\usepackage{graphicx}
\usepackage[normalem]{ulem}

\usepackage{graphicx}
\usepackage{dcolumn}
\usepackage{bm}

\renewcommand{\d}[1]{\textrm{d}#1} 
\def\lsim{\mathrel{\raise.3ex\hbox{$<$\kern-.75em\lower1ex\hbox{$\sim$}}}}
\def\gsim{\mathrel{\raise.3ex\hbox{$>$\kern-.75em\lower1ex\hbox{$\sim$}}}}

\begin{document}

\preprint{APS/123-QED}

\title{Searching for dark matter annihilation in the Sun with the IceCube Upgrade}

\author{Dan Hooper\orcidlink{0000-0001-8837-4127}}
    \email{dwhooper@wisc.edu}
\affiliation{
    Department of Physics, Wisconsin IceCube Particle Astrophysics Center, University of Wisconsin, \\
    Madison, Wisconsin 53706, USA
}
\author{Fabrizio Vassallo\orcidlink{0000-0002-0541-5606}}
    \email{fevassallo@wisc.edu}
\affiliation{
    Department of Physics, Wisconsin IceCube Particle Astrophysics Center, University of Wisconsin, \\
    Madison, Wisconsin 53706, USA
}

\begin{abstract}
    The IceCube Upgrade will provide unprecedented sensitivity to dark matter particles annihilating in the core of the Sun. For dark matter candidates with spin-dependent couplings to nuclei and that annihilate significantly to tau leptons or neutrinos, we find that the IceCube Upgrade will be capable of testing parameter space that is beyond the reach of existing direct detection experiments. After calculating the sensitivity of the IceCube Upgrade to dark matter annihilation in the Sun, we explore dark matter models that could be tested by this experiment, identifying two classes of scenarios as promising targets for such searches.
\end{abstract}

\maketitle


\section{Introduction}

Many strategies have been pursued in an effort to detect the products of dark matter annihilation, including searches for gamma-rays from the Galactic Center~\cite{Daylan:2014rsa,Hooper:2011ti,Hooper:2010mq,Goodenough:2009gk,Fermi-LAT:2015sau,Fermi-LAT:2017opo}, as well as from the Galactic halo~\cite{Fermi-LAT:2012pls,Fermi-LAT:2013thd}, and from satellite galaxies of the Milky Way~\cite{McDaniel:2023bju,DiMauro:2022hue}. Searches for dark matter annihilation products in the cosmic-ray spectrum have also been carried out, focusing largely on positrons~\cite{Bergstrom:2013jra}, antiprotons~\cite{Calore:2022stf,Cuoco:2016eej,Cui:2016ppb,Cholis:2019ejx}, and antinuclei~\cite{DeLaTorreLuque:2024htu,Winkler:2022zdu,Cholis:2020twh}. Searches have also been conducted for neutrinos that could be produced through dark matter annihilation in the core of the Sun~\cite{IceCube:2021xzo,IceCube:2019lus,IceCube:2016dgk,IceCube:2016yoy,ANTARES:2016xuh,Super-Kamiokande:2015xms,IceCube:2012ugg,IceCube:2011aj,IceCube:2009iyf,Super-Kamiokande:2004pou,Berlin:2024lwe,Maity:2023rez,Krishna:2025ncv}.

Seven new strings of optical modules will be deployed this year within the existing volume of the IceCube Neutrino Observatory, constituting what is known as the IceCube Upgrade~\cite{Ishihara:2019aao}. The smaller spacing between these modules, both along a given string and between the strings, will allow the upgraded IceCube detector to observe neutrino-induced muon tracks with energies as low as a few GeV, well below the threshold of the existing observatory. Among other scientific goals~\cite{IceCube-Gen2:2019fet,IceCube:2023ins,Rongen:2021rgc,Baur:2019jwm}, the IceCube Upgrade will have unprecedented sensitivity to dark matter particles annihilating in the core of the Sun, in particular for dark matter candidates that annihilate significantly to tau leptons or directly to neutrinos.


Stringent constraints have already been placed on dark matter candidates that could be captured by the Sun and annihilate within its core. Direct detection experiments, in particular, have placed strong limits on the dark matter's elastic scattering cross section with nuclei~\cite{LZ:2024zvo,XENON:2025vwd,PICO:2019vsc,CRESST:2022lqw,CRESST:2019jnq}, limiting the rate at which dark matter particles could scatter with and be captured by the Sun. The annihilation cross section of dark matter particles has also been constrained by measurements of the angular power spectrum of the cosmic microwave background (CMB)~\cite{Calabrese:2025mza,Planck:2018vyg}. In light of these and other constraints, it is not necessarily trivial to identify viable dark matter models that could be within the reach of the IceCube Upgrade. 

In this paper, we estimate the sensitivity of the planned IceCube Upgrade to dark matter particles annihilating in the Sun. After taking into account constraints from direct detection experiments, we find that dark matter could produce a detectable flux of neutrinos from the Sun only if these particles couple to nuclei predominantly though spin-dependent couplings and annihilate significantly to tau leptons or neutrinos. We take a bottom-up approach, considering in generality the range of possible dark matter candidates that could produce a sizable flux of high-energy neutrinos from the Sun, and identify two classes of dark matter models that could potentially be probed by the planned IceCube Upgrade.

\section{Dark Matter Capture and Annihilation in the Sun}

It has long been appreciated that dark matter particles could potentially scatter with and become gravitationally bound to the Sun~\cite{1985ApJ...299.1001K,Press:1985ug,Griest:1986yu,Gould:1987ju,Gould:1989hm}. If enough dark matter particles accumulate in the Sun's core, this could lead to a high annihilation rate and to a potentially observable flux of high-energy neutrinos~\cite{Silk:1985ax,Srednicki:1986vj,Kamionkowski:1991nj,Barger:2001ur,Flacke:2009eu,Schuster:2009au,Batell:2009zp}. The number of dark matter particles that are gravitationally bound to the Sun, $N_{\chi}$, evolves according to the sum of their capture, annihilation, and evaporation rates,
\begin{align}
\label{eq:differential}
    \frac{\d{N_\chi}}{\d{t}} &= \Gamma_{\text{cap}} - 2 \Gamma_{\text{ann}} -\Gamma_{\rm evap}.
\end{align}
%
%
The capture rate is determined by integrating over the volume of the Sun and over the dark matter's velocity distribution~\cite{Widmark_2017, Catena_2015, Gould_1987ii}, 
\begin{align}
    \Gamma_{\text{cap}} = 4\pi \int_0^{R_\odot}  r^2 dr \int_0^\infty  \frac{f(u)}{u} \, w \, \Omega_v(w) \, du,
\end{align}
where $f(u)$ is the dark matter's velocity distribution, $w = \sqrt{u^2 + v(r)^2}$, $v(r)$ is the local escape velocity from the Sun, and $\Omega_{v}(w)$ is the probability that a given dark matter particle will elastically scatter with a nucleus and be left with a velocity below the Sun's escape velocity. This latter quantity is given by
\begin{align}
    \Omega_v (w) = \sum_i n_i \, w \, \Theta \left(\alpha_i - \frac{u^2}{w^2}\right) \int_{m_\chi u^2/2}^{\alpha_i m_\chi w^2/2}  \frac{d\sigma_i}{dE} (w^2, q^2) \, dE,
\end{align}
where the sum is performed over the nuclear species of interest, each with a mass, $m_i$, and a local number density, $n_i$. The mass of the dark matter particle is denoted by $m_{\chi}$, while $\alpha_i = 4m_\chi m_i / (m_\chi + m_i)^2$, $d\sigma_i/dE$ is the differential elastic scattering cross section, and $E = q^2/2m_i$ is the nuclear recoil energy. We take the velocities of the dark matter to follow a Maxwell-Boltzmann distribution with a dispersion of $\bar{v}=270 \, {\rm km/s}$ and boosted by 220 km/s with respect to the Sun. We normalize this distribution to a density of 0.4 GeV/cm$^3$. Note that the dark matter density and velocity distribution near the Sun are well approximated by their values in the local halo \cite{Sivertsson:2012qj, Gould:1991}.

Numerically, this capture rate can be approximately expressed as follows~\cite{Gould:1991hx}:
\begin{align}
\Gamma_{\rm cap} &\approx 1.7 \times 10^{21} \, {\rm s}^{-1} \times \bigg(\frac{100 \, {\rm GeV}}{m_{\chi}}\bigg) \bigg(\frac{\rho_{\chi}}{0.4 \, {\rm GeV}/{\rm cm}^3}\bigg) \bigg(\frac{270 \, {\rm km/s}}{\bar{v}}\bigg) \\
&\times \bigg[\bigg(\frac{\sigma_{\chi p}}{10^{-42}\, {\rm cm}^2}\bigg)S(m_{\chi}/m_p) + 1.1 \, \bigg(\frac{\sigma_{\chi {\rm He}}}{16 \times 10^{-42}\, {\rm cm}^2}\bigg)S(m_{\chi}/4m_p)\bigg], \nonumber
\end{align}
where $S(x)= [A(x)^{3/2}/(1+A(x)^{3/2})]^{2/3}$ and $A(x)=3x \,v_{\rm esc}^2/2(x-1)^2 \bar{v}^2$.

Once gravitationally captured, the dark matter particles undergo a series of subsequent scatters which cause them to lose additional energy and to accumulate near the Sun's core. The effective volume of the region that these particles occupy, $V_{\rm eff}$, is found by matching the temperature of the Sun's core to the gravitational potential energy of an individual dark matter particle~\cite{Gould:1987ir,Griest:1986yu}. The total dark matter annihilation rate in the Sun can be found by integrating the local annihilation rate, $n_{\chi}^2 \langle \sigma v \rangle$, over the volume of the Sun, resulting in
\begin{align}
\Gamma_{\rm ann} = 2\pi \int_0^{R_\odot}  n_\chi^2 \langle \sigma v \rangle \, r^2 \,dr \approx \frac{N_{\chi}^2 \,\langle \sigma v \rangle}{2 V_{\rm eff}}, \label{eq:annihilation_rate}
\end{align}
where in the last step we have taken $\langle \sigma v \rangle$ to be independent of velocity (i.e., an $s$-wave process).

Very light dark matter particles can evaporate from the Sun at a significant rate. The rate for this process is approximately given by~\cite{Gould:1987ju,Griest:1986yu}
\begin{align}
\frac{\Gamma_{\rm evap}}{N_{\chi}} \approx  
2 \times 10^{-8} \, {\rm s}^{-1} \times 
10^{-3.5 (m_{\chi}/{\rm GeV})} \,\bigg(\frac{\sigma_{\chi p}}{10^{-42} \, {\rm cm}^2}\bigg).
\end{align}
Notice that the effects of evaporation are very sensitive to the precise value of the dark matter's mass. For the case of $\sigma_{\chi p} \sim 10^{-42} \, {\rm cm}^2$, the timescale for evaporation  is less than that the age of the Solar System only for $m_{\chi} \lsim 3 \, {\rm GeV}$. For dark matter particles that are potentially detectable by the IceCube Upgrade, evaporation is expected to have a negligible impact. 

Neglecting evaporation and using Eq.~\ref{eq:annihilation_rate}, the solution to Eq.~\ref{eq:differential} corresponds to the following rate of dark matter annihilation in the Sun today:
\begin{align}
\Gamma_{\rm ann} = \frac{\Gamma_{\rm cap}}{2} \,\tanh^2 \bigg( \frac{t_{\odot} \sqrt{2 \Gamma_{\rm cap} \Gamma_{\rm ann}}}{N_\chi} \bigg),
\end{align}
where $t_{\odot} \approx 4.5 \times 10^9 \, {\rm yr}$ is the age of the Solar System. Thus, for $t_{\odot} \gg N_\chi/\sqrt{2 \Gamma_{\rm cap}\Gamma_{\rm ann}}$, equilibrium will be reached between the rates of capture and annihilation. For $m_{\chi} \gg m_p$, capture-annihilation equilibrium is obtained so long as the following condition is satisfied:
\begin{align}
    \sigma_{\chi p} \gtrsim 10^{-41} \, {\rm cm}^2 \times \bigg(\frac{2.2\times 10^{-26} \, {\rm cm}^3/{\rm s}}{\langle \sigma v \rangle}\bigg) \bigg(\frac{100 \, {\rm GeV}}{m_{\chi}}\bigg)^{1/2}.
\end{align}
Note that in our calculations, we do not assume that capture-annihilation equilibrium is achieved.

With the capture and annihilation rates in hand, we can calculate the differential fluxes of muon neutrinos and antineutrinos that reach Earth after being produced through dark matter annihilation in the Sun~\cite{Catena_2015, Edsjo_1997}:
\begin{align}
    \frac{d \phi_{\nu_{\mu}}}{dE_{\nu_{\mu}}} &= \frac{1}{3} \frac{\Gamma_{\text{ann}}}{4 \pi D^2} \bigg(\frac{dN_{\nu_e}}{dE_{\nu_e}}+\frac{dN_{\nu_{\mu}}}{dE_{\nu_{\mu}}}+\frac{dN_{\nu_{\tau}}}{dE_{\nu_{\tau}}}\bigg)_{\rm Inj} \\
    \frac{d \phi_{\bar{\nu}_{\mu}}}{dE_{\bar{\nu}_{\mu}}} &= \frac{1}{2}\frac{\Gamma_{\text{ann}}}{4 \pi D^2} \bigg(\frac{dN_{\bar{\nu}_{\mu}}}{dE_{\bar{\nu}_{\mu}}}+\frac{dN_{\bar{\nu}_{\tau}}}{dE_{\bar{\nu}_{\tau}}}\bigg)_{\rm Inj},\nonumber
\end{align}
where $D$ is the Earth-Sun distance, and the terms in brackets are the initial spectra of neutrinos or antineutrinos which are produced per annihilation. Note that the factors of 1/3 and 1/2 in the front of these expressions account for the effects of neutrino oscillations.\footnote{Due to the Mikheyev–Smirnov–Wolfenstein (MSW) effect, electron neutrinos contribute to the flux of muon neutrinos that reach Earth. In contrast, electron antineutrino oscillations are MSW suppressed.} In calculating the injected spectra of neutrinos and antineutrinos, we have considered dark matter annihilations to $\nu \bar{\nu}$, $\tau^+ \tau^-$, $b\bar{b}$, and $c\bar{c}$~\cite{Jungman:1994jr} (see also, Refs.~\cite{Hooper:2008cf,Fitzpatrick:2010em}). Any muons or charged pions that are produced in the core of the Sun will lose the vast majority of their energy before decaying and thus do not contribute to the resulting high-energy neutrino flux. For annihilations to $\tau^+ \tau^-$, we include neutrinos produced through the following decays: $\tau \rightarrow \mu \nu \nu$, $e\nu\nu$, $\pi \nu$, $K\nu$, $\pi \pi \nu$, and $\pi \pi \pi \nu$. For the hadronic annihilation channels, only semileptonic decays contribute, $q \rightarrow q' l \nu $. Finally, we approximate the impact of neutrino absorption in the Sun by including an additional factor of $e^{-E_{\nu}/E_{\rm abs}}$, where $E_{\rm abs} = 130 \, {\rm GeV}$ for neutrinos and $E_{\rm abs} = 200 \, {\rm GeV}$ for antineutrinos~\cite{Crotty:2002mv}.

\section{The IceCube Upgrade}

The IceCube Upgrade, which is scheduled to be deployed in 2025–26, will consist of seven new strings carrying approximately 700 optical sensors, densely embedded near the bottom-center of the existing IceCube Neutrino Observatory at the South Pole. These upgrade strings will make up a 20 meter (horizontal) by 3 meter (vertical) grid of detectors, located 2150 to 2425 meters below the surface of the Antarctic ice~\cite{Ishihara:2019aao}.

We estimate the rate of neutrino-induced muon tracks by the following:
\begin{align}
\Gamma_{\mu} &\approx N_A \int \int \frac{d\phi_{\nu_{\mu}}}{dE_{\nu_{\mu}}}(E_{\nu_{\mu}}) \, \frac{d\sigma_{\nu}}{dy}(E_{\nu_{\mu}}, y) \, D_{\mu}(E_{\mu}) \, A_{\mu, {\rm eff}} \, dE_{\nu_{\mu}} \, dy \\
&+ N_A \int \int \frac{d\phi_{\bar{\nu}_{\mu}}}{dE_{\bar{\nu}_{\mu}}}(E_{\bar{\nu}_{\mu}}) \,\frac{d\sigma_{\bar{\nu}}}{dy}(E_{\bar{\nu}_{\mu}}, y) \,D_{\mu}(E_{\mu}) \, A_{\mu, {\rm eff}} \,dE_{\nu_{\mu}} \, dy,\nonumber
\end{align}
where $N_A$ is number density of nucleons in ice, $d\phi_{\nu_{\mu}}/dE_{\nu_{\mu}}$ is the flux of neutrinos that reaches the detector, $d\sigma_{\nu}/dy$ is the differential cross section for neutrino-nucleon charged-current scattering~\cite{Gandhi:1995tf}, and $E_{\mu} \equiv (1-y) E_{\nu_{\rm}}$. For the dimensions of the IceCube Upgrade, we adopt a width and height of 80 m and 275 m, respectively, corresponding to an effective area of $A_{\mu, {\rm eff}} = 22,000 \,{\rm m}^2$. For the effective depth of the detector, we take $D_{\mu} = 60 \, {\rm m} + R_{\mu}$, where the physical depth of the detector is 80 m and $R_{\mu} \approx 2380 \, {\rm m} \times \ln[(2+0.0042E_{\mu})/2.0168]$ is the distance that a muon travels before dropping below the detection threshold of 4 GeV~\cite{Dutta:2000hh}. For muons with an initial energy of less than 4 GeV, we take $D_{\mu}=0$.


After calculating the signal rate, we compare this to the rate of background events from atmospheric neutrinos~\cite{Super-Kamiokande:2015qek}. Integrated over a $25^{\circ}$ angular radius around the Sun, this background results in a rate of 590.8 events per year. At the 2$\sigma$ confidence level, we project that after 10 years of observation, the IceCube upgrade will be sensitive to signal rates larger than $\Gamma_{\mu} > 2 \sqrt{590.8\times 10}/10 \approx 15.4 \, {\rm yr}^{-1}$.

\begin{figure}
    \centering
    \includegraphics[width=0.7\linewidth]{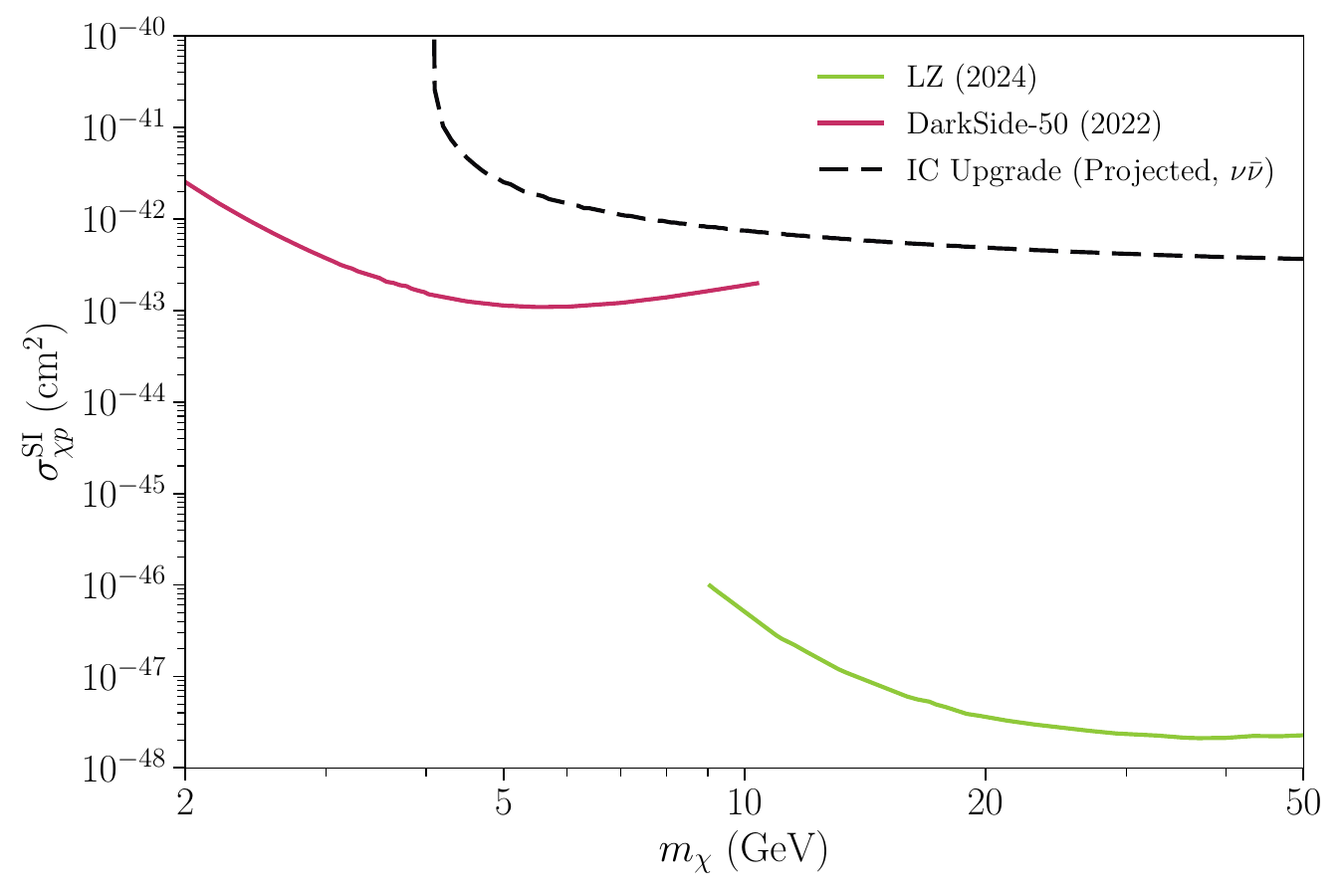}
    \caption{The projected sensitivity of the IceCube Upgrade to dark matter particles annihilating in the Sun after 10 years of observation, compared to existing constraints from the direct detection experiments LZ~\cite{LZ:2024zvo} and DarkSide-50~\cite{DarkSide-50:2022qzh}. Here, we have considered the case of spin-independent (SI) scattering and annihilation to $\chi \chi \rightarrow \nu \bar{\nu}$ (equally to all flavors) with a cross section of $\langle \sigma v \rangle =2.2 \times 10^{-26} \,{\rm cm}^3/{\rm s}$. From this comparison, we see that the IceCube Upgrade is not expected to be sensitive to dark matter particles that are captured in the Sun through spin-independent scattering, even for the optimistic case of annihilations that proceed directly to neutrino pairs.}
    \label{fig:SI}
\end{figure}

In Fig.~\ref{fig:SI}, we show the projected sensitivity of the IceCube Upgrade to dark matter annihilating in the Sun to $\chi \chi \rightarrow \nu \bar{\nu}$ (equally to all flavors), for the case of spin-independent elastic scattering (such that $\sigma_{\chi N} \propto A^2$, where $A$ is the atomic mass of the nuclear target). We compare this projected sensitivity to the current sensitivity of the direct detection experiments LZ~\cite{LZ:2024zvo} and DarkSide-50~\cite{DarkSide-50:2022qzh}. From this comparison, we see that the IceCube Upgrade is not expected to be sensitive to dark matter particles that are captured in the Sun through spin-independent scattering, even for the optimistic case of annihilations which proceed directly to neutrino pairs. 

\begin{figure}
    \centering
    \includegraphics[width=0.7\linewidth]{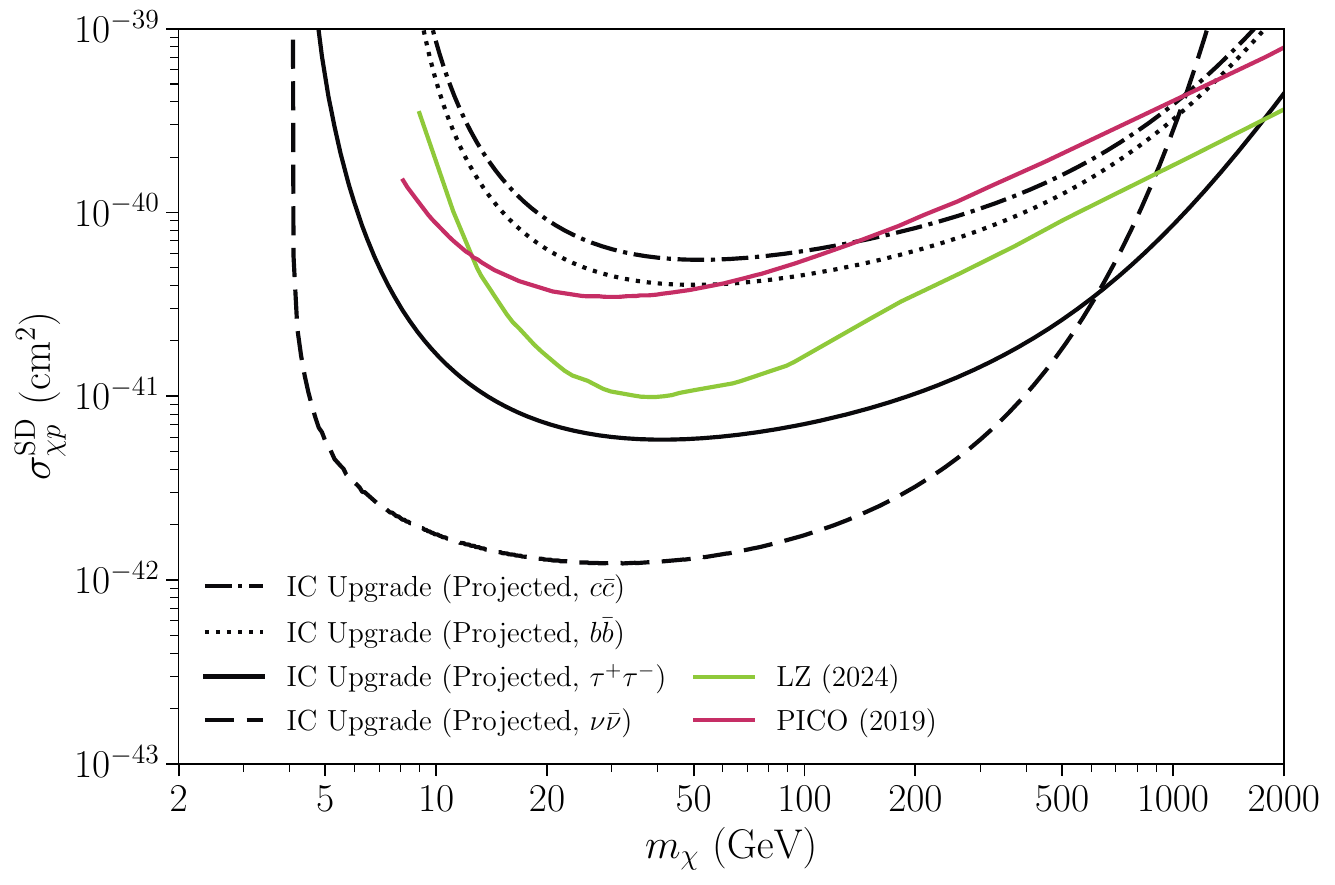}
    \caption{The projected sensitivity of the IceCube Upgrade to dark matter particles annihilating in the Sun after 10 years of observation, compared to the existing constraints from the direct detection experiments LZ~\cite{LZ:2024zvo} and PICO~\cite{PICO:2019vsc}. Here, we have considered spin-dependent (SD) scattering and annihilations to $\chi \chi \rightarrow \nu \bar{\nu}$ (equally to all flavors), $\tau^+ \tau^-$, $c\bar{c}$, or $b\bar{b}$, in each case with a cross section of $\langle \sigma v \rangle =2.2 \times 10^{-26} \,{\rm cm}^3/{\rm s}$.}
    \label{fig:SD}
\end{figure}

In light of this result, we will focus throughout the remainder of this study on dark matter candidates that are captured in the Sun predominantly through spin-dependent scattering, such that $\sigma_{\chi N} \propto J(J+1)$, where $J$ is the spin of the nuclear target. In Fig.~\ref{fig:SD}, we plot the projected sensitivity of the IceCube Upgrade to dark matter with spin-dependent couplings for annihilations to $\chi \chi \rightarrow \nu\bar{\nu}$, $\tau^+ \tau^-$, $b\bar{b}$, or $c\bar{c}$. Again, we compare these projections to the existing constraints from direct detection experiments, in this case being LZ~\cite{LZ:2024zvo} and PICO~\cite{PICO:2019vsc}. In contrast to the spin-independent case, we find that the IceCube Upgrade will provide the most stringent constraints on dark matter's spin-dependent scattering cross section for the masses in the range of $m_{\chi} \sim 5$–$1700 \, {\rm GeV}$ for annihilations to $\tau^+ \tau^-$, and $m_{\chi} \sim 4$–$600 \, {\rm GeV}$ for annihilations to $\nu \bar{\nu}$.

\begin{figure}
    \centering
    \includegraphics[width=0.495\linewidth]{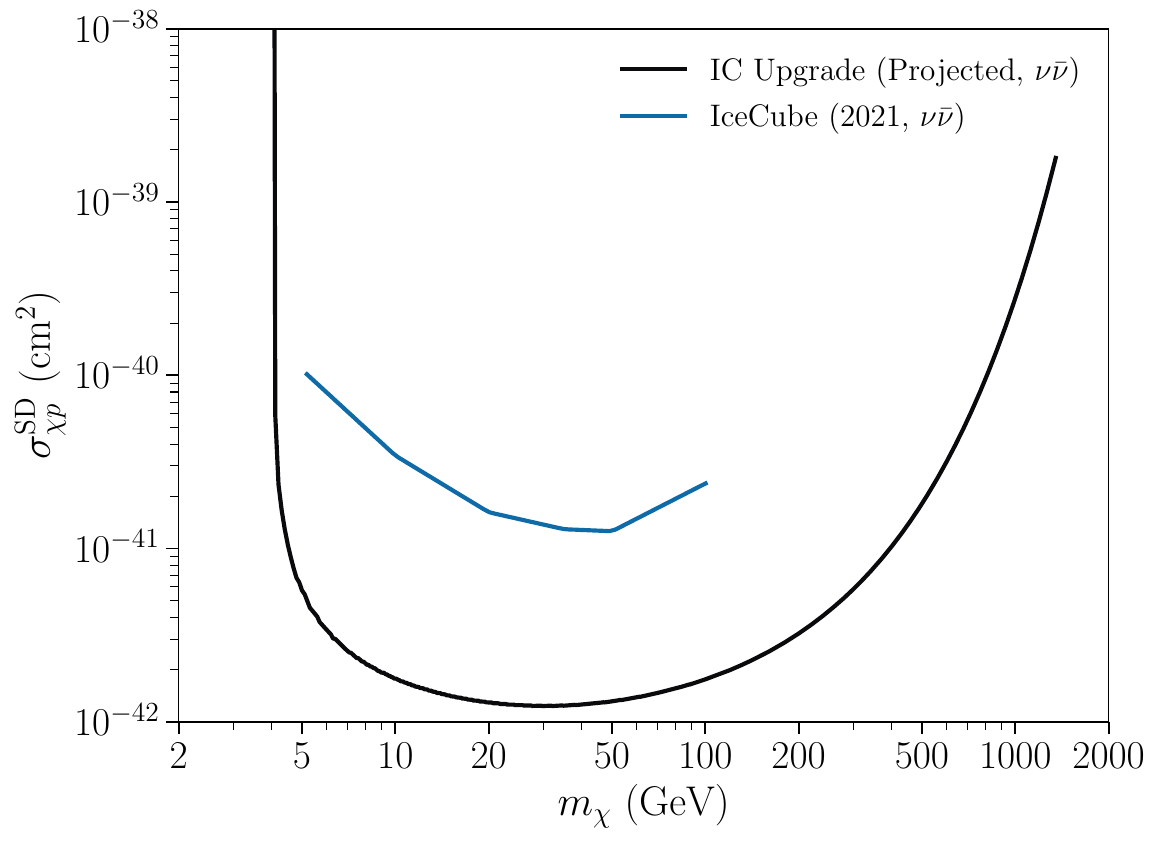}
    \includegraphics[width=0.495\linewidth]{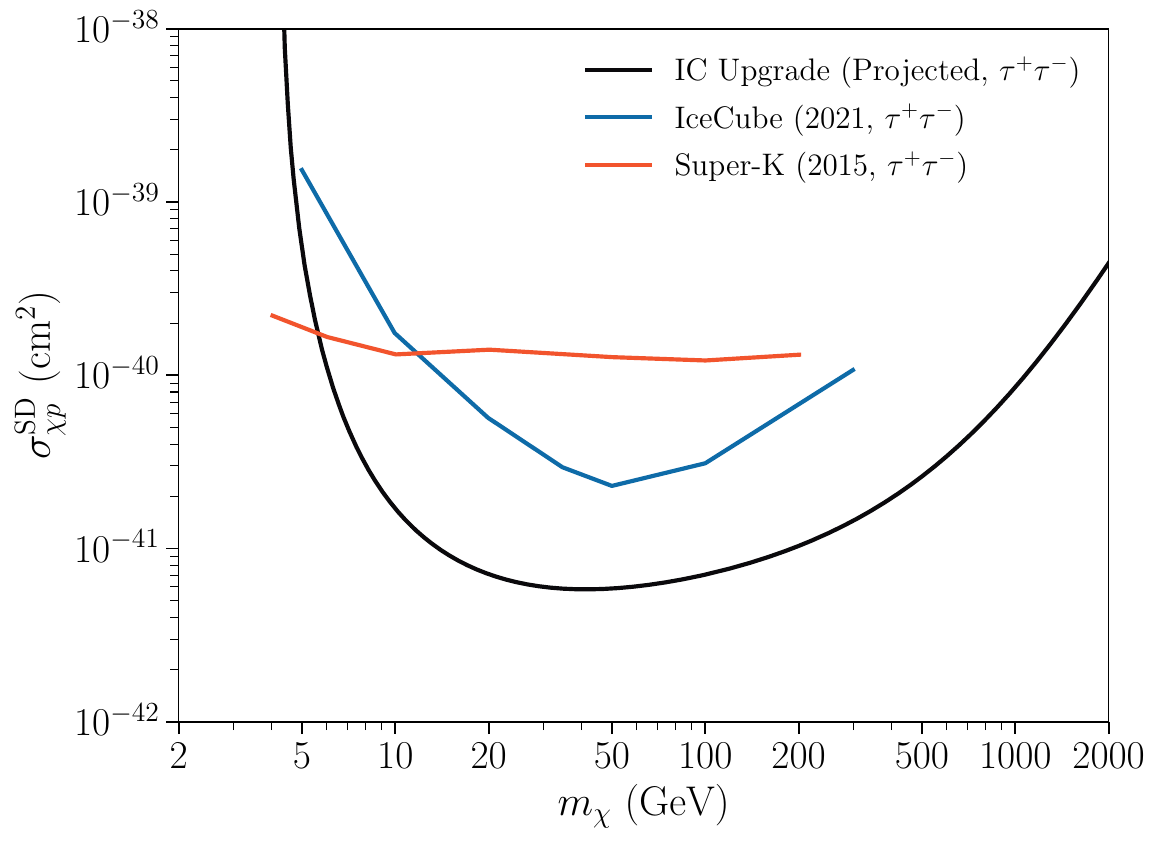} \\
    \includegraphics[width=0.495\linewidth]{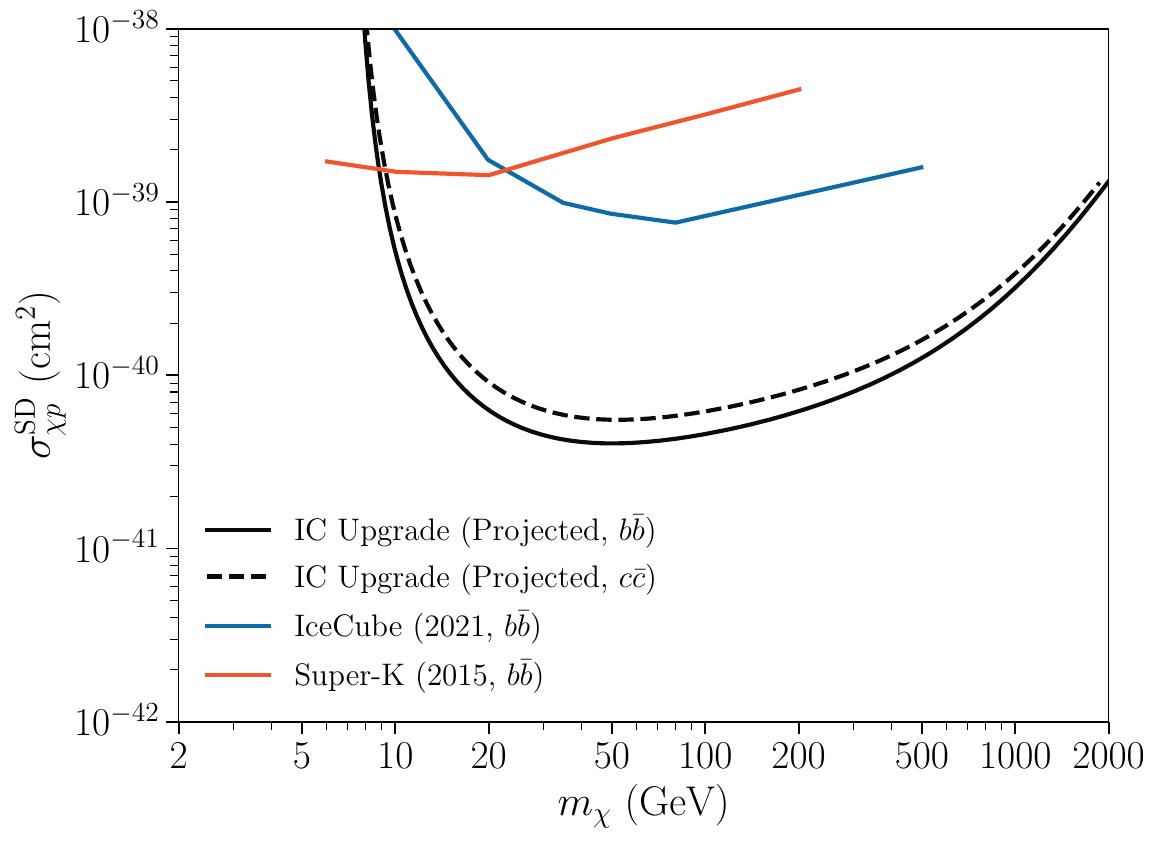}
    \caption{A comparison of the projected sensitivity of the IceCube Upgrade (after 10 years of observation) to neutrinos from dark matter annihilations in the Sun to existing constraints from the IceCube DeepCore~\cite{IceCube:2021xzo} and Super-Kamokande~\cite{Super-Kamiokande:2015xms} detectors. The low threshold and large volume of the IceCube Upgrade will enable this detector to offer unprecedented sensitivity to neutrinos from dark matter annihilation in the Sun for a wide range of masses.}
    \label{fig:ID}
\end{figure}

In Fig.~\ref{fig:ID}, we compare the projected sensitivity of the IceCube Upgrade to the existing constraints from the IceCube DeepCore~\cite{IceCube:2021xzo} and Super-Kamokande~\cite{Super-Kamiokande:2015xms} detectors (for the projected sensitivity of Hyper-Kamiokande, see Ref.~\cite{Bell_2021}). The low threshold and large volume of the IceCube Upgrade will enable this detector to offer unprecedented sensitivity to neutrinos produced through dark matter annihilations in the Sun. We note that when the IceCube Collaboration presents their determination of the projected sensitivity of the IceCube Upgrade to dark matter annihilation in the Sun, based on a detailed Monte Carlo study, those results should be taken to supersede the estimates presented in this section of this paper. 

\section{Dark Matter Models}

In this section, we set out to identify those dark matter models that will fall within the projected reach of the IceCube Upgrade while being consistent with existing constraints from direct detection experiments~\cite{LZ:2024zvo,XENON:2025vwd,PICO:2019vsc,CRESST:2022lqw,CRESST:2019jnq}, accelerators~\cite{CMS:2022usq,CMS:2016gsl,ATLAS:2018qto,CMS:2016cfx,CMS:2018ipm,ATLAS:2019erb}, and measurements of the CMB~\cite{Calabrese:2025mza,Planck:2018vyg}. Throughout this analysis, we require that the dark matter candidate freezes out of equilibrium in the early Universe with a thermal relic abundance equal to the measured density of dark matter.

For the reasons described in the previous section, we will focus here on dark matter candidates that scatter with nuclei predominantly through spin-dependent couplings. To facilitate a high solar capture rate, we further require that this spin-dependent scattering cross section is not suppressed at low velocities (such as in those models in which $\sigma^{\rm SD}_{\chi p}$ scales with powers of the momentum transferred, or transverse velocity). As enumerated in Ref.~\cite{Berlin:2014tja}, there are only two types of interactions between dark matter and quarks that satisfy these constraints. These are, 1) fermionic (Dirac or Majorana) dark matter that scatters with nuclei through the $t$-channel exchange of a spin-one mediator with axial ($\gamma^{\mu} \gamma^5$) couplings to both dark matter and quarks, and 2) dark matter in the form of a Majorana fermion that scatters with nuclei through the $s$-channel exchange of a spin-zero mediator with a scalar-pseudoscalar ($1 \pm \gamma^5$) coupling to dark matter and either up or down quarks. For a complete list of such interactions and their resulting spin-independent and spin-dependent scattering cross sections, see Ref.~\cite{Berlin:2014tja}.

\subsection{Axial models}
\label{sec:axial}

The following Lagrangian describes the interactions of the dark matter with Standard Model fermions in this model:
\begin{equation}
\mathcal{L} \supset \bigg[ g_{\chi} \bar{\chi} \gamma^\mu   \gamma^5  \chi + \sum_f g_f \bar{f} \gamma^\mu  \gamma^5  f \bigg] Z'_\mu,\end{equation}
where our dark matter candidate, $\chi$, is either a Dirac or Majorana fermion. The mediator in this model is a spin-one particle, $Z'$, with couplings parametrized by $g_{\chi}$ and $g_{f}$. 

This leads to a dark matter annihilation cross section that is given by 
\begin{align}
    \sigma & = \sum_f \frac{n_f g_{\chi}^2 g_f^2}{12 \pi s \left[ \left(s-m_{Z'} ^2\right)^2 + m_{Z'}^2 \Gamma_{Z'}^2 \right]} \sqrt{\frac{1-4 m_f^2/s}{1-4 m_{\chi }^2/s}} \\
    & \times 
    \left[4 m^2_{\chi}m^2_f \bigg(7-\frac{6s}{m_{Z'}^2}+\frac{3s^2}{m^4_{Z'}}\bigg) -4m_{\chi}^2 s + s(s-4m^2_f) \right], \nonumber 
\end{align}
where $n_f =1 (3)$ for leptons (quarks), and $s$ is the standard Mandelstam variable. Expanding this in powers of velocity, we arrive at
\begin{align}
\langle\sigma v \rangle &\approx \sum_f \frac{n_f g_{\chi}^2 g_f^2 m^2_f \sqrt{1-m^2_f/m^2_{\chi}}}{2 \pi (m_{Z'}^2-4 m_{\chi}^2)^2} + \frac{n_f v^2 g_{\chi}^2 g_f^2 }{48 \pi  (m_{Z'}^2-4m^2_{\chi})^2 \sqrt{1-m_f^2/m^2_{\chi}}} \Bigg[
-\frac{72 m^4_f}{m^2_{Z'}}+\frac{17 m^4_{f}}{m_{\chi}^2}  \nonumber \\
&\,\,\,\,\,\,\,\,\,\,\,\,\,\,\,\,\,\,\,\,\,\,+\frac{144 m^2_{\chi} m_f^4 }{m^4_{Z'}}+ \frac{48 m^2_{\chi} m^2_f}{m_{Z'}^2} -22 m^2_f -\frac{96 m^4_{\chi} m^2_f}{m^4_{Z'}} +8 m^2_{\chi}
 \Bigg].
\end{align}
%

\begin{figure}
    \centering
    \includegraphics[width=0.495\linewidth]{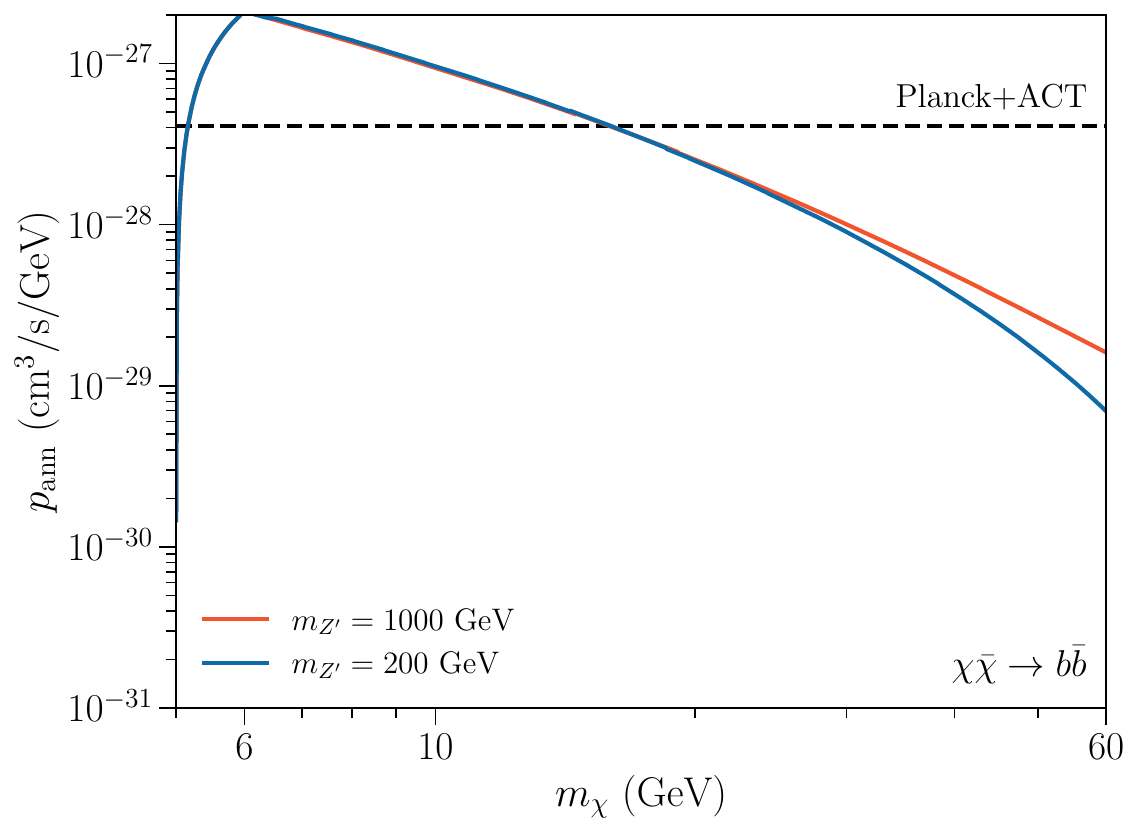}
    \includegraphics[width=0.495\linewidth]{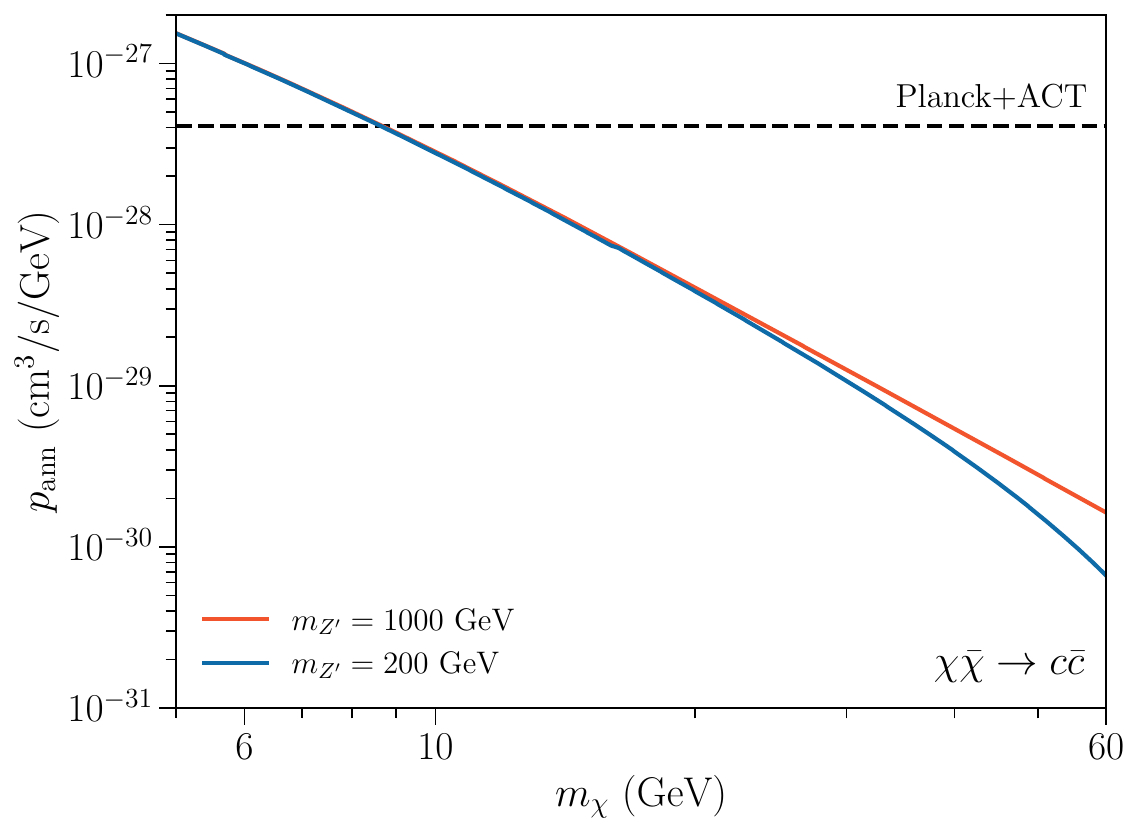} \\
    \includegraphics[width=0.495\linewidth]{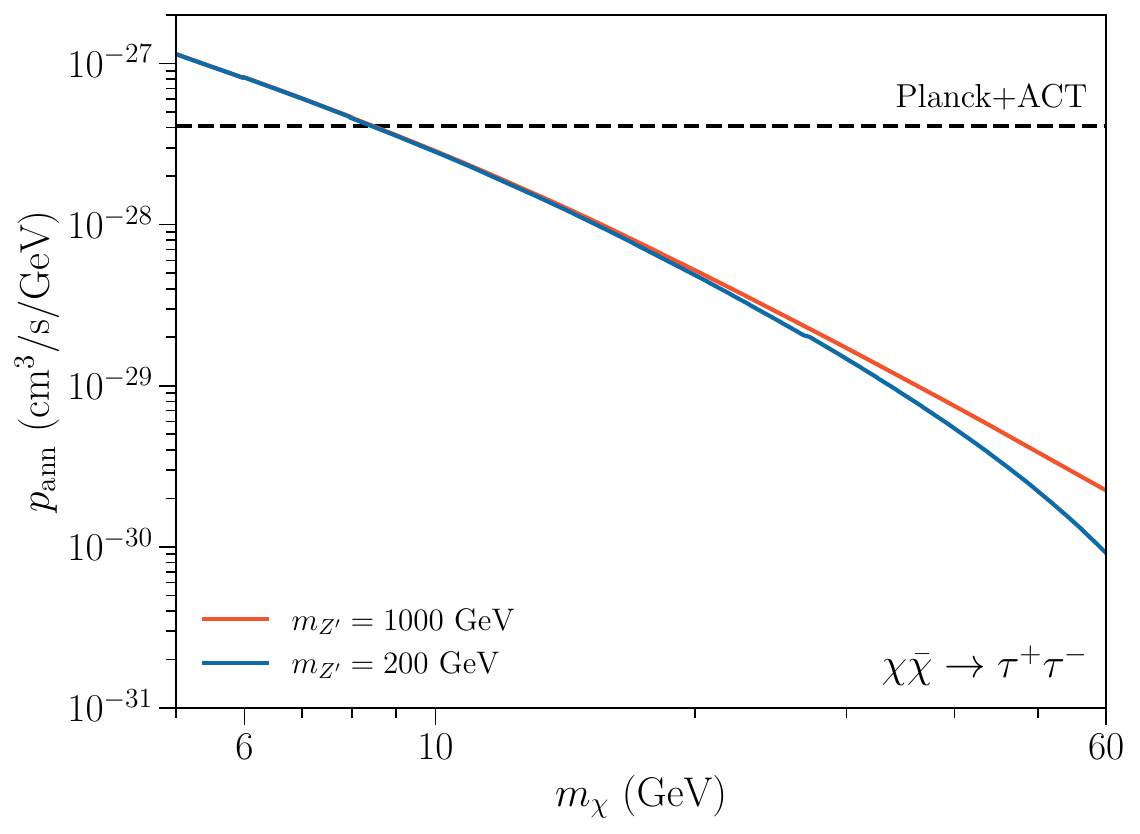}
    \caption{The value of $p_{\rm ann} \equiv f_{\rm eff} \langle \sigma v \rangle_{v \rightarrow 0} /m_{\chi}$ predicted in the axial dark matter model described in Sec.~\ref{sec:axial}, where $f_{\rm eff}$ is the fraction of the energy released in these annihilations that is transferred into the intergalactic medium during the relevant epoch, $z \approx 600$~\cite{2012PhRvD..85d3522F,Slatyer:2015jla}.}
    \label{fig:pann}
\end{figure}

For each choice of masses, we set the product of the couplings to obtain a thermal relic abundance equal to the measured cosmological dark matter density, $\Omega_{\chi}h^2 = 0.12$~\cite{Planck:2018vyg}. For these parameter values, we then calculate the dark matter annihilation rate during the era of recombination and compare this to the constraints from the Atacama Cosmology Telescope (ACT) and other CMB telescopes~\cite{Calabrese:2025mza}. In Fig.~\ref{fig:pann}, we plot the quantity, $p_{\rm ann} \equiv f_{\rm eff} \langle \sigma v \rangle_{v \rightarrow 0} /m_{\chi}$, that is predicted in this model, where $f_{\rm eff}$ is the fraction of the energy released in these annihilations that is transferred into the intergalactic medium during the relevant epoch, $z \approx 600$~\cite{2012PhRvD..85d3522F,Slatyer:2015jla}. From this comparison, we find that measurements of the CMB rule out dark matter masses lighter than 17 GeV, 9 GeV, and 9 GeV for annihilations purely to $b\bar{b}$, $c\bar{c}$, or $\tau^+ \tau^-$, respectively. 

The dark matter's elastic scattering cross section with nucleons is purely spin-dependent in this model, and is given by
\begin{align}
\sigma_{\chi (p, n)} &\approx \frac{3 \mu^2_{\chi (p,n)} g^2_{\chi}}{\pi m^4_{Z'}}  \, \bigg( \sum_{q=u,d,s} g_{q}\Delta^{(p, n)}_q\bigg)^2 \\
& \approx 1.6 \times 10^{-41} \, {\rm cm}^2 \times \bigg(\frac{g_{\chi}}{1}\bigg)^2 \bigg(\frac{300 \,\rm GeV}{m_{Z'}}\bigg)^4 \bigg(\frac{\sum g_{q}\Delta^{(p, n)}_q}{0.02}\bigg)^2,\nonumber
\end{align}
where $\mu_{\chi (p,n)}$ is the reduced mass of the system, and $\Delta_u^{(p)}=\Delta_d^{(n)}=0.84$, $\Delta_u^{(n)}=\Delta_d^{(p)}=-0.43$, and $\Delta_s^{(p)}=\Delta_s^{(n)}=-0.09$~\cite{Cheng:2012qr}.

\begin{figure}
    \centering
\includegraphics[width=0.495\linewidth]{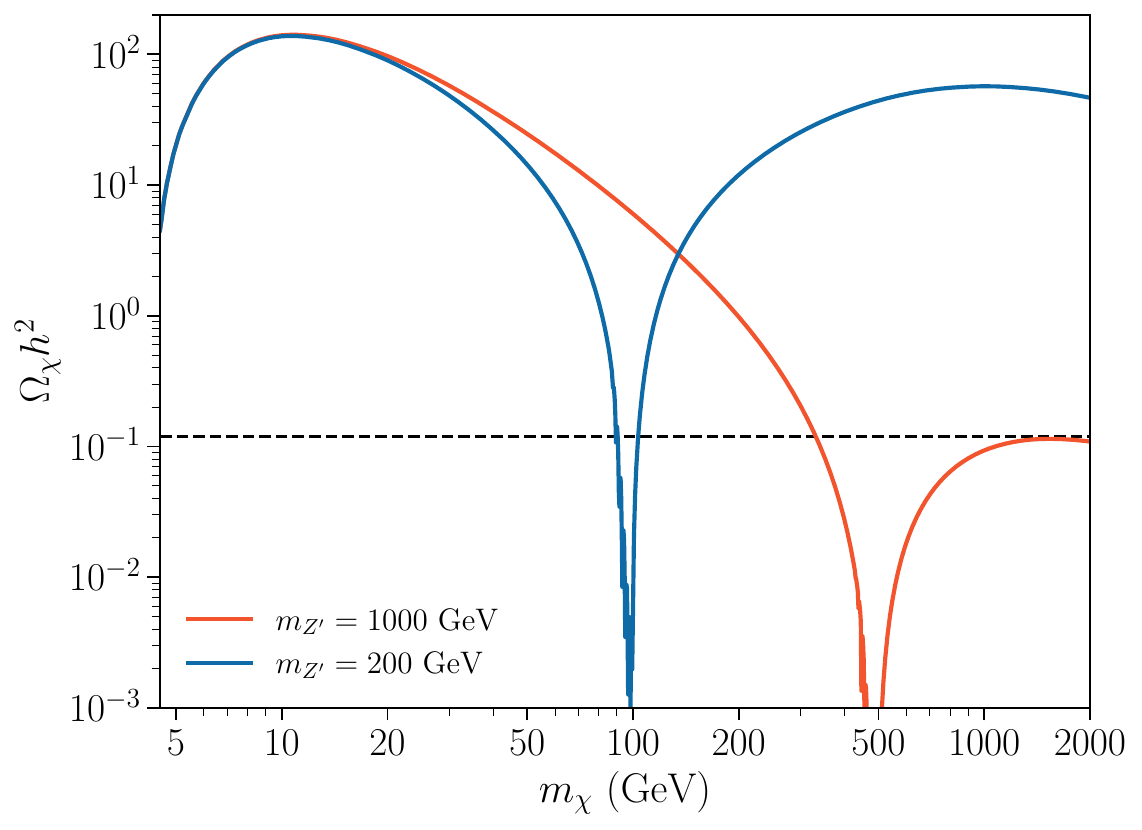} 
\includegraphics[width=0.495\linewidth]{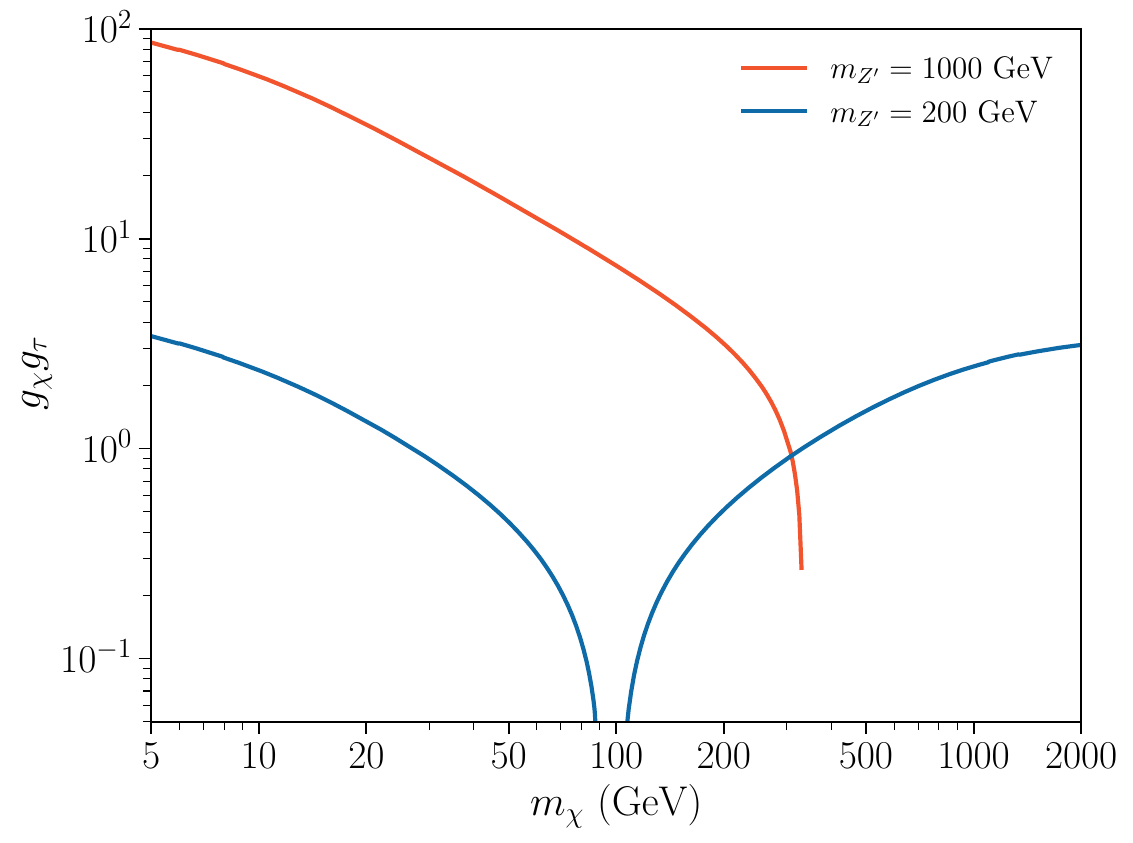} 
    \caption{Left frame: The value of thermal relic abundance that is obtained in the axial model, considering only couplings to dark matter and to up and down quarks, and after setting these couplings to obtain a spin-dependent cross section that is equal to the projected sensitivity of the IceCube Upgrade as shown in Fig.~\ref{fig:SD} (for the case of annihilations to $\tau^+ \tau^-$). Right frame: The product of the couplings, $g_{\chi} g_{\tau}$, that leads to a thermal relic abundance of $\Omega_{\chi} h^2 =0.12$, for the same couplings to quarks that were adopted in the left frame.}
    \label{fig:axial}
\end{figure}

In the left frame of Fig.~\ref{fig:axial}, we plot the value of thermal relic abundance that is obtained in this model, considering only couplings to dark matter and to up and down quarks, and after setting these couplings to obtain a spin-dependent cross section that is equal to the projected sensitivity of the IceCube Upgrade as shown in Fig.~\ref{fig:SD} (for the case of annihilations to $\tau^+ \tau^-$). This figure illustrates that for $m_{Z'}=1 \, {\rm TeV}$ and $m_{\chi} \gsim 300 \, {\rm GeV}$, for example, the dark matter particles annihilate too efficiently in the early Universe to make up the entire dark matter abundance. For lighter values of $m_{\chi}$, other couplings (such as those which result in annihilations to $\tau^+ \tau^-$) could enhance the annihilation rate, leading to an acceptable thermal relic abundance. In the right frame of this figure, we show the product of the couplings, $g_{\chi} g_{\tau}$, which leads to a thermal relic abundance of $\Omega_{\chi} h^2 =0.12$, where $g_{\tau}$ is the coupling of the mediator to charged leptons and for the same couplings to up and down quarks that were adopted in the left frame. 

For a wide range of dark matter masses, there exist combinations of $m_{Z'}$, $g_{\chi}$, $g_{u,d}$, and $g_{\tau}$ which yield the measured dark matter abundance and that predict a neutrino flux from the Sun that is within the projected reach of the IceCube Upgrade. Across much of this parameter space, the hadronic couplings can be sufficiently small to evade dijet searches at the LHC and other colliders~\cite{CMS:2022usq,CMS:2016gsl,ATLAS:2018qto}, for perturbative values of the dark matter coupling, $g_{\chi} \lsim 4 \pi$. Searches for dielectron or dimuon resonances~\cite{CMS:2016cfx,CMS:2018ipm,ATLAS:2019erb} would rule out this class of models, however, unless the mediator's coupling to taus is significantly larger than those to electrons or muons, $g_{\tau} \gg g_{e}, g_{\mu}$.

Putting this information together, we conclude that a fermionic dark matter candidate with interactions mediated by a spin-one particle with axial couplings could produce a flux of neutrinos from the Sun that is within the projected sensitivity of the IceCube Upgrade. This requires the dark matter to annihilate significantly to tau leptons and for the mediator to possess sizable couplings to light quarks, $g_{u,d} \sim (0.01-0.2) \times (g_{\chi}/1) (m_{Z'}/300 \, {\rm GeV})^2$.

%

\subsection{Scalar-pseudoscalar models}
\label{sec:SP}

As a second class of models, we will consider the following Lagrangian:
\begin{align}
  \mathcal{L} \supset   \lambda_{\chi q} \bigg(\bar{\chi} (1 + \gamma^5 ) q A_Q +  \bar{q} ( 1 -  \gamma^5 ) \chi A_Q^\dagger\bigg), 
\end{align}
where $\lambda_{\chi q}$ is the coupling between our Majorana fermion dark matter candidate, $\chi$, the spin-zero mediator, $A_Q$, and a light quark, either $u$, $d$, or $s$. 
Notice that the mediator in this model, $A_Q$, carries both QCD color and fractional electric charge. For such a particle to evade constraints from the LHC, we must require $m_{A_Q} \gsim 2 \, {\rm TeV}$, in close analogy to the constraints placed on up and down squarks~\cite{CMS:2023xlp,CMS:2019zmd,ATLAS:2024lda,ATLAS:2021twp}. 

Expanding in powers of velocity, the dark matter's annihilation cross section in this model is given by 
\begin{align}
    \sigma v  \approx \frac{3 \lambda_{\chi q}^4 m_q^2 \sqrt{1 - m_q^2/m_\chi^2}}{2(m_{A_Q}^2 + m_\chi^2)^2} + \frac{ \lambda_{\chi q}^4 (   m_{A_Q}^4   +  m_\chi^4 ) v^2}{ \pi m_\chi^4 (m_{A_Q}^2 + m_\chi^2)^4 \sqrt{1 - m_q^2/m_\chi^2}}.
\end{align}
Notice that the $v^2$ term in this expansion (the $p$-wave contribution) dominates over the velocity-independent term, leading to very weak constraints from measurements of the CMB.

The spin-dependent scattering cross section in this model is given by~\cite{Berlin:2014tja}
\begin{align}
\sigma_{\chi (p, n)} &\approx \frac{3 \mu^2_{\chi (p,n)}} {4\pi m^4_A}  \bigg( \sum_{q=u,d,s} \lambda^2_{\chi q}\Delta^{(p, n)}_q\bigg)^2 \\
& \approx 1.2 \times 10^{-41} \, {\rm cm}^2 \times \bigg(\frac{3 \,\rm TeV}{m_A}\bigg)^4 \bigg(\frac{\sum \lambda^2_{\chi q}\Delta^{(p, n)}_q}{2}\bigg)^2.\nonumber
\end{align}
From this expression, we see that to generate an elastic scattering that is large enough to produce an observable flux of neutrinos from the Sun, we need either $\lambda_{\chi u} \sim 1.5 \times (m_A/3 \, {\rm TeV})$ or $\lambda_{\chi d} \sim 2.2 \times (m_A/3 \, {\rm TeV})$ . 

\begin{figure}
    \centering
    \includegraphics[width=0.495\linewidth]{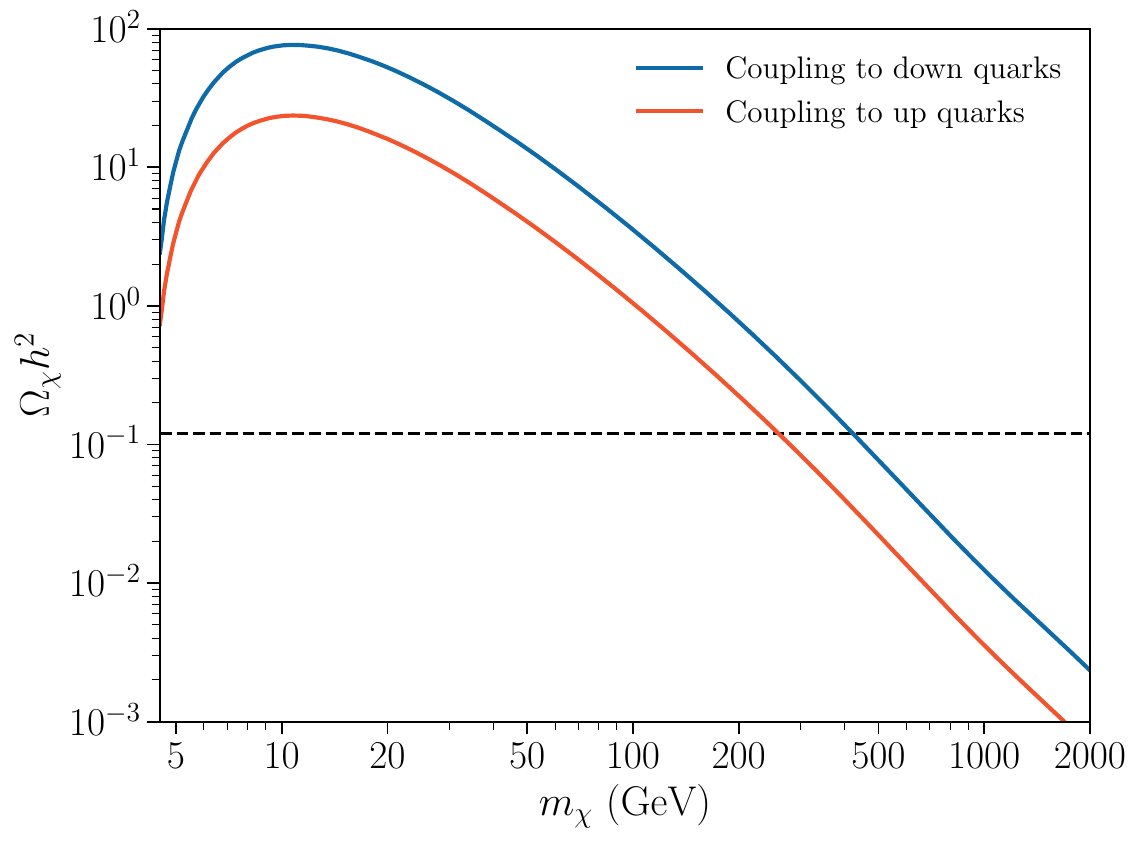}
    \caption{The value of thermal relic abundance that we obtain in the scalar-pseudoscalar model, for the case of a 2 TeV mediator with couplings to dark matter and to either up or down quarks, after setting this coupling to the value required to obtain a spin-dependent cross section equal to the projected sensitivity of the IceCube Upgrade as shown in Fig.~\ref{fig:SD} (for the case of annihilations to $\tau^+ \tau^-$).}
    \label{fig:SP}
\end{figure}

In Fig.~\ref{fig:SP}, we plot the value of thermal relic abundance that is obtained in this model, considering only couplings to dark matter and to either up or down quarks, and after setting this coupling such that the spin-dependent cross section is equal to the projected sensitivity of the IceCube Upgrade as shown in Fig.~\ref{fig:SD} (for the case of annihilations to $\tau^+ \tau^-$). For $m_{\chi} \gsim 400 \, {\rm GeV}$, the dark matter particles annihilate too efficiently in the early Universe to make up the entire dark matter abundance. For lighter values of $m_{\chi}$, other annihilation channels could enhance the annihilation rate, leading to an acceptable thermal relic abundance, and providing a means by which a flux of high-energy neutrinos could be generated in the Sun. 


Putting this information together, we conclude that dark matter in the form of a Majorana fermion with interactions that are mediated by a spin-zero particle with a scalar-pseudoscalar coupling to light quarks could lead to a large capture rate in the Sun. If there exists another coupling that allows for the efficient annihilation of these particles into tau leptons, this could result in a flux of neutrinos from the Sun that is within the projected sensitivity of the IceCube Upgrade. 

\section{Summary and Discussion}

Seven new strings of optical modules are scheduled to be deployed later this year within the bottom-center volume of the existing IceCube Neutrino Observatory at the South Pole. The resulting IceCube Upgrade will provide unprecedented sensitivity to dark matter particles annihilating in the core of the Sun, in particular in the case of dark matter candidates that annihilate to tau leptons or directly to neutrinos.

For dark matter particles to be efficiently captured through elastic scattering with protons or nuclei in the Sun without exceeding constraints from direct detection experiments, this scattering must occur through unsuppressed spin-dependent couplings. This requirement significantly limits the range of dark matter models to which the IceCube Upgrade could potentially be sensitive. In particular, this condition is satisfied for only two types of interactions. In the first case, the dark matter is a fermion that scatters with nuclei through the $t$-channel exchange of a spin-one mediator with axial ($\gamma^{\mu} \gamma^5$) couplings to both the dark matter and to light quarks. In the second case, the dark matter is a Majorana fermion that scatters with nuclei through the $s$-channel exchange of a spin-zero mediator with a scalar-pseudoscalar ($1 \pm \gamma^5$) coupling to the dark matter and to either up or down quarks. 

For each of these classes of dark matter models, we have identified regions of parameter space that are consistent with all existing constraints, including those from direct detection experiments, accelerator-based experiments, and measurements of the CMB. In the axial case, the spin-one mediator must have couplings to both light quarks and tau leptons, requiring exotic fermions to cancel the corresponding gauge anomalies (for a review, see Ref.~\cite{Langacker:2008yv}). In the scalar-pseudoscalar case, a second interaction channel is required to facilitate the annihilations of dark matter particles to either tau leptons or neutrinos.

\begin{acknowledgments}
    This work has been supported by the Office of the Vice Chancellor for Research at the University of Wisconsin-Madison, with funding from the Wisconsin Alumni Research Foundation.
\end{acknowledgments}

\bibliography{bibliography}

\end{document}